\documentclass[twocolumn,aps,pra,showpacs]{revtex4}

\usepackage{epsfig}
\usepackage{graphicx}
\usepackage{amssymb}
\usepackage{multirow}
\usepackage{textcomp}
\usepackage{enumitem}

\newcommand {\ket}[1] {|{#1}\rangle}

\newcommand {\bra}[1] {\langle{#1}|}

\newcommand {\norm} [1] {\parallel #1 \parallel}

\newcommand {\beq} {\begin{equation}}
\newcommand {\eeq} {\end{equation}}
\newcommand {\mrm}{\mathrm}

\begin{document}


\title{Fundamental Quantitative Security In Quantum Key Generation}


\author{Horace P. Yuen\\Department of Elecrical Engineering and Computer Science,\\Department of
Physics and Astronomy,\\Northwestern University, Evanston
IL, 60208,\\Email:yuen@eecs.northwestern.edu}

\date{\today}
\begin{abstract}
We analyze the fundamental security significance of the quantitative criteria on the final generated
key $K$ in quantum key generation including the quantum criterion $d$, the attacker's mutual
information on $K$, and the statistical
distance between her distribution on $K$ and the uniform distribution. For operational significance
a criterion has to produce guarantee on the attacker's probability of correctly
estimating
some portions of $K$ from her measurement, in particular her maximum probability
of identifying the whole $K$. We distinguish between the raw security of $K$ when the
attacker just gets at $K$ before it is used in a cryptographic context and its composition
security when the attacker may gain further information during its actual use to help getting at
$K$. We compare both of these securities of $K$ to those obtainable from conventional key expansion
with a symmetric
key cipher. It is pointed out that a common belief in the superior security of a quantum generated
$K$ is based on an incorrect interpretation of $d$ which cannot be true, and
the security significance of $d$ is uncertain. Generally, the QKD key $K$ has no composition
security guarantee and its raw security guarantee from concrete protocols is worse than that of
conventional ciphers. Furthermore, for both raw and composition
security there is an exponential catch up problem that would make it difficult to quantitatively
improve the security of $K$ in a realistic protocol. Some possible ways to deal with the situation
are
suggested.
\end{abstract}
\maketitle
\section*{I. Introduction And Summary}
Physical cryptography, the use of physical effects in addition to purely mathematical artifice for
fast reliable cryptographic functions, has received considerable recent attention. For key
generation, variously called key distribution, key expansion, key exchange or key agreement, the use
of classical noise was first proposed [1]-[3] while the use of information-disturbance tradeoff in
BB84 type protocols are the well known quantum key distribution (QKD) schemes [4]-[5]. Other
quantum schemes that dispense with intrusion level estimation have been developed on the basis of
incompatible quantum measurements in the KCQ (keyed communication in quantum noise) approach
[6]-[9]. The fundamental merit of these physical cryptographic schemes is that the so-called
information theoretic (IT) security is possible, in contrast to the expansion of a
master key to session keys or key agreement from public-key protocols with
security based on computational complexity. It is the purpose of this paper to demonstrate in
detail that, contrary to widespread perception and belief,  (i)  it is not clear how strong IT
security can even be obtained in principle from QKD; (ii) the security guarantee that can
be experimentally obtained thus far is quite inadequate.  Some assessment and suggestion will be
made on the current situation.  Note that we are not at all concerned with appropriate system
modeling or device imperfection issues, but rather just with the fundamental quantitatively
achievable security in a concrete protocol with realistic parameters, assuming perfect devices and
applicable system model.
\par There are \textit{two} main reasons for this present unsatisfactory state of affairs: The
problem of quantitative security criteria and the security of the final generated key $K$ during
actual
use in a cryptography scheme \textbf{\textemdash} the so-called composability problem.
We will explain them by comparing a perfect fresh key $K^{\mathsf{p}}$ and a session key $K'$
generated from a pseudorandom number generator (PRNG) to the key \textit{K} generated from BB84 type
protocols [10]. The comparison between $K'$ and $K$ is meaningful because a shared secret key
between the users is also needed in KCQ, and in QKD for message authentication, during the key
generation process. The keys obtainable from KCQ or classical noise-based protocols would share all
or
some of
the problems associated with \textit{K}.
\par
A perfect key $K^{\mathsf{p}}$ has a uniform probability distribution $p(K^{\mathsf{p}}) = U$ to the
attacker (Eve).
For $K^{\mathsf{p}}$ consisting of $n$ bits, there are $2^n$ possible values for $K^{\mathsf{p}}$
each with a
probability $U_i=2^{-n}$. Furthermore, any piece of additional information Eve may obtain in the
actual use of $K^{\mathsf{p}}$ in any cryptographic scheme should not give Eve more information
about
$K^{\mathsf{p}}$.
For example, when $K^{\mathsf{p}}$ is used in one-time pad encryption of some $n$ data bits and
$\textit{m}<\textit{n}$ bits are revealed in a known-plaintext attack (KPA), Eve still would have a
uniform distribution with respect to the remaining $n-m$ bits in $K^{\mathsf{p}}$. We would
call the pre-use quantitative security of \textit{K} its ``\textit{raw security}'' and the in-use
security its
``\textit{composition security}''. Composition security guarantee
is needed even when the raw security is adequate. In this paper we would deal
with composition security mainly on the problem of the extent any bit of $K$ remains secret when
some other part of $K$ is known to Eve.
For such problem whether $K$ is perfect for a classical protocol reduces to the question whether
$p(K) = U,$ where $p(K)$ is Eve's probability distribution on $K$ obtained through her attack with
possible further information gained during the actual use of $K$.
The situation is far more complicated in the quantum case due to the possibility of quantum memory.
It is possible to obtain a perfect key in practice when
a secure method (but perhaps clumsy such as hand delivery) is used to deliver a randomly
picked key
string to be shared between two users Alice and Bob. If one generates the key by a
public-key technique such as RSA, its security is based entirely on (presumed) computational
complexity, i.e., it is (presumably) practically impossible for Eve to determine \textit{K} due to
the lack of an efficient algorithm even though it is possible in principle.  In contrast,
$K^{\mathsf{p}}$ has IT security that is not changed by Eve's computational power.

\par One often hears that the generated key \textit{K} in QKD has IT security, by itself a rather
misleading statement because the \textit{K} obtained in
physical cryptography is \textit{never} perfect just in the sense of $p(K)=U$. A mistaken claim is
maintained in the literature[5],[11]-[12] that $p(K)=U$ with a high probability if one uses
the criterion $d$ to be discussed later. It is one of the main purposes of this paper to dispel this
misconception. In contrast to a QKD generated key $K$, a fresh key when available can
typically be taken to be perfect in standard cryptography as indicated above. A more
precise security claim on $K$ is that it is sufficiently close to perfect. But then the crucial
issue of quantitative security criteria arises for measuring the closeness, and in particular
\textit{why} any specific achievable quantitative value is adequately secure in a given
application. This problem does \textit{not} arise in a standard fresh key for which $p(K)=U$.

\par There is no fundamental guarantee of IT security from ``randomness test'' given Eve's
knowledge of the key generation process. The \textit{proper security criterion} on $K$ is the set of
probabilities $p(\tilde{K})$ of Eve's correct estimates on all the possible subsets $\tilde{K}$ of
$K$ which she could derive from her possible $p(K)$. Common single-number criterion such as her
mutual information on $K$ has no empirical operational security significance in itself and merely
expresses a constraint on Eve's possible $p(K)$. For operational guarantee one would need to
translate such a criterion into guarantees on $p(\tilde{K})$ which would be carried out in this
paper.

\par To avoid possible confusion one may distinguish the following three different logical
situations:
\begin{enumerate}
 \item[(a)] A proof of security has been obtained that works against all possible attacks with high
probability.
\item[(b)] A specific attack has been found that breaches security with high probability.
\item[(c)] Security level unknown for various possible attacks.
\end{enumerate}
In this paper we are not talking about case (b). Instead, it is pointed out that case (a) has not
been established and case (c) is the current situation contrary to the claims in the QKD literature.

\par For comparison to the case of \textit{K}, we first describe the
raw and composition security for the key $K'$ obtained by PRNG in standard (conventional)
cryptography. In
typical "key expansion" scheme a master key $K^{\mathsf{m}}$ is fed through a PRNG to generate a key
$K'$ with many more bits that $K^{\mathsf{m}}$, $\left| K^{\mathsf{m}} \right|< \left|K' \right|$.
Different segments of $K'$ are then used as different session keys in various uses in order to
reduce the total number of perfect key bits otherwise needed.  On the raw security of $K'$, the
Shannon measure of information or Eve's entropy on $K'$, $H_E\left(K' \right)$, is often employed.
The well known Shannon limit [3],[9],[13]-[14] says

 \begin{equation}\label{1}
H_E\left(K'\right) \leqslant \left| K^{\mathsf{m}} \right|
 \end{equation} If $\tilde{K}'$ is any subset (subsequence) of  $K'$, it is possible that
$H_E(\tilde{K}')=|\tilde{K}'|$ similar to the case of a uniform $\tilde{K}'$ even under (1)
when $|\tilde{K}'|\leq |K^{\mathsf{m}}|$, but no other $\tilde{K}'$
can
have a uniform distribution since that would violate (1). Thus, with $H_E(K')$ taken as the
measure
of raw security, there is IT security for such session keys in standard cryptography \textit{also}.
It is
just that their quantitative level may be far from perfect.

\par There are other important measures of quantitative IT security:  Eve's maximum probability
$p_1(K)$ of determining the whole key $K$, her maximum probabilities $p_1(\tilde{K})$ of
determining various subsets $\tilde{K}$ or $K$, and these are the ones with operational
significance. In addition to $H_E$, there is another common single-number measure used in
quantifying the randomness of a bit sequence, namely the statistical distance $\delta(p(K),U)\equiv
\delta_E(K)$ between Eve's probability distribution of $K$ and $U$. Generally,
from different attacks on a physical cryptosystem and from different measured ciphertexts, Eve would
obtain \textit{different} probability
distribution $p(K)$ on $K$, which would determine the above quantities and whatever other measure
one may employ. The significant point and a main difficulty is that no single number, be it
$H_E, p_1$  or $\delta_E$, could capture the full security picture in physical key generation with
even just one probability distribution $p(K)$ for Eve.

\par
For raw security, many PRNG including those given just by a (maximum length) linear feedback shift
registrar (LFSR) with perfect seedkey $K^{\mathsf{m}}$ has the following behavior [15],
 \begin{equation}\label{2}
p_1(K')=2^{-|K^{\textsf{m}}|},\qquad  p_1(\tilde{K}')=2^{-|\tilde{K}'|}
 \end{equation} where $\tilde{K}'$ is a subsequence of $K'$ up to $|K^{\mathsf{m}}|$ consecutive
bits.
This is
very
favorable compared to the $K$ that can be generated by QKD as will be seen in the next paragraph.
However, such $K'$ has no IT composition security.  Specifically, if  $K'$ is used in "one-time pad"
form, i.e., the PRNG is used as an additive stream cipher with  $K'$ as running key, then a KPA with
$|K^{\mathsf{m}}|$ known consecutive data bits would lead to a unique determination of
$K^{\mathsf{m}}$ for the usual
nondegenerate ciphers including LFSRs in their common cipher configurations.  The knowledge of
$K^{\mathsf{m}}$ would then allow the complete determination of $K'$ [14]-[15]. The situation is
similar
when $K^{\mathsf{p}}$ is used in a conventional symmetric-key block cipher such as AES.

\par For $K$ obtained from QKD, the most commonly used criterion for raw security is
$I_E/|K|=1-H_E(K)/|K|$, Eve's information per bit on $K$ from her attack. Under the quantitative
security
$I_E/|K|\leq2^{-l}$ , it was shown [9, section IIIB] the possibility is not ruled out that Eve may
obtain a correct estimate of $K$  and \textit{m}-bits subsets $\tilde{K}$  of $K$ with
probabilities, for $l < |K|$,
 \begin{equation}\label{3}
p_1(K)\sim2^{-l},\qquad  p_1(\tilde{K}')\sim \frac{n}{m}\:2^{-l}
 \end{equation} Such possibility arises because the bits in $K$ are not statistically independent to
Eve. In BB84 they have been correlated from error correction and privacy amplification as well as
from Eve's joint attack. For concrete
protocols that can be experimentally developed thus far [16],
$l\sim10$ for $|K| > 10^3$
which means a disastrous breach of security with $p_1(K)\sim10^{-3}$ is not ruled out. In any event,
even $l\sim50$ is rather unfavorable compared to (2) where $|K^{\mathsf{m}}|>100$ is typical in most
conventional ciphers.

\par Another useful criterion is the statistical distance $\delta_E(K)$ between
$p(K)$ and $U$.
For  $\delta_E(K) = 2^{-l}$ with $l\leq|K|$, the possibility remains that [9, App. B]
 	 \begin{equation}\label{4}
p_1(K)\sim2^{-l},\qquad  p_1(\tilde{K})\sim2^{-l}
 \end{equation} It should be noted that a sufficiently small $p_1(K)$ guarantee is
evidently \textit{necessary} for meaningful security guarantee, see Section IV. It is apparent from
(3)-(4) that according to the criterion $I_{E}/|K| \leq \epsilon$ and $\delta_{E} \leq \epsilon$,
$K$
is only nearly uniform when $\epsilon = 2^{-\lambda|K|}$ with $\lambda \sim 1$, a condition that
appears impossible to achieve by a realistic protocol with a significant key generation rate. For
the criterion $I_E/|K|$, QKD does much better than PRNG which is responsible for
the better composition security of $K$ as
will be shown in this paper.

\par On the composition security of $K$ it has been shown [12] that when Eve retains her probe, for
$I_E/|K|\sim2^{-l}$ it may be possible for her to tell the $(l+1)$th bit of $K$ knowing $l$ of them
from, say, a KPA on the of use of $K$ as one-time pad. To overcome this problem, the use of a
different criterion $d$,
 	 \begin{equation}\label{5}
d\equiv\frac{1}{2}\|\rho_{KE} -\rho_U\otimes\rho_E\|_{1}\leq\epsilon
 \end{equation} was suggested and developed [11], [17].  The claim is that under (5) the users
would get a perfect key $K$ with probability at least $1-\epsilon$. This is an incorrect
interpretation of (5) as has been mentioned [9, App. B]. Given this prevailing misconception
we will give a detailed discussion in section IIIC and bring out the point instead that $p(K)$ is
actually
\textit{never} given by $U$ for $d = \epsilon >0$ and that $d\leq\epsilon$ has \textit{no} clear
raw or composition security significance. In any event, theoretical estimates [18]-[19] give
$\epsilon = 10^{-5}$ for various large $|K|$, corresponding at best to (4) with $l\sim17$.

\par As will be detailed in the paper, the following problem situation obtains on the security
guarantee of the generated key
$K$ in
concrete QKD protocols:
	\begin{enumerate}[label=(\roman{*})]
   \item The raw security of $K$ is worse than that of a LFSR for the probabilities  $p_1(K)$
of identify the whole $K$ by an attacker and  $p_1(\tilde{K})$ for many smaller subsets $\tilde{K}$
of $K$.
   \item There is no composition security guarantee for $K$; an exponential decrease of the
accessible information may only lead to a linear decrease of the number of compromised key bits
while the situation is
unknown for $\delta_{E}$ or $d$.
 \end{enumerate}

\par In the following we will flesh out these points and discuss their implications on the
development of physical cryptography.  In section II we will discuss the raw security of the
generated key  $K$. In section III the composition security
of $K$ will be treated which shows the lack of such guarantee thus far for both specific
attack scenarios and in general. In Section IV we discuss the relevance of rigorous security proofs
and the importance of actual numerical values. Some suggestions on possible future development of
physical cryptography is outlined in section V.
\section*{II. RAW SECURITY OF THE GENERATED KEY}
\par We will first review the raw security of a running key  $K'$  generated from a master key in
standard cryptography for later comparison with the key generated from QKD. Recall that by the
\textit{raw security} of $K$ we mean the quantitative security level of $K$ against attacks during
the key generation process without any additional information that Eve may obtain during its actual
cryptographic use. For a key obtained from
public-key technique such as RSA, there is no IT security at all since the key can be determined by
Eve if she has sufficient computational power.  Thus, we would not further discuss public-key
schemes in this paper which deals only with IT security. For the usual symmetric-key
ciphers there is IT security for $K'$ and it is better than that
of concrete realistic QKD schemes. As we shall see, this arises as a result of the detailed
quantitative behavior of the security measures that have been adopted to describe the
quantitative security. We will begin with a discussion of the security measures.

\subsection*{IIA.  Security Criterion and Security Parameter}

\par The attacker's optimal probabilities $p_1(\tilde{K})$ of correctly estimating $\tilde{K}$, any
subset of $K$, constitute the operationally meaningful criterion on the security of $K$. In
conventional cryptography the closeness of $p_{1}(\tilde{K})$ with $p_{1}(\tilde{K}^{\mathsf{p}})$
for all $\tilde{K}$ is called \textit{semantic security} [20], whether polynomial complexity is
included [21] on an attack algorithm or not [22]. A security parameter $\epsilon$ may be introduced
to measure the deviation of such semantic security from a perfect key $K^\mathsf{p}$, $\epsilon = 0$
corresponding to the perfect case.

\par Such semantic security guarantee is difficult to obtain and to prove. Simpler
criterion is often employed instead. However, theoretical constructs such as the attacker's mutual
information $I_{E}(K)$ on $K$ has no operational meaning by itself. In the context of
communications, these information theoretic quantities derive their empirical significance via the
Shannon Coding
Theorems through which they are related to operational quantities [23]. In the context of
cryptography, operational significance could only be obtained from the security
guarantee on the various $p_1(\tilde{K})$ implied by such single-number criterion. This is what we
would spell out in this paper for the common criteria
$I_{E}(K) / |K|$ and $\delta_{E}$.

\subsection*{IIB.  Conventional Key Expansion Raw IT Security}
\par Consider the running key  $K'$ generated from a PRNG with master key $K^{\mathsf{m}}$  , with
number of bits  $|K'|>|K^{\mathsf{m}}|$. Segments of $|K'|$ could be used as session keys in
different applications. It can also be used as a running key, i.e., in one-time pad form as an
additive stream cipher. Let $X_n$
 be the \textit{n}-bit data random variable, we use lower case $\textbf{x}_n$ to
denote a
specific \textit{n}-sequence from $X_n$. An additive stream cipher would have
output $Y_n$ with $\textbf{y}_n=\textbf{x}_n\oplus \textbf{k}_n$, where $\textbf{k}_n$ is a
specific value of the n-bit $K_{n}'$. (We add the subscript n on occasion for clarity.) The general
Shannon limit for a
standard cipher is
given by (1). It holds for any cipher that satisfies unique decryption $H(X_n|Y_n,
K^{\mathsf{m}})=0$, thus
covers also `random ciphers' with randomized encryption [13]-[14].

To derive (1) the so-called Kerckhoff's assumption [14] is sometimes invoked which says Eve knows
everything about the cipher except the seedkey $K^{\mathsf{m}}$. Such an assumption is not actually
needed so
long as all the information Alice and Bob need to share but which Eve does not have can be
quantified by a proper key $K^{\mathsf{m}}$.   The ciphertext $Y_n$ is also assumed to be available
to Eve,
evidently a most common situation in reality.

\par We will discuss several quantitative criteria of security. First Eve has only one probability
distribution on the possible $K_{n}'$ for nonrandom
nondegenerate ciphers.  All standard ciphers are nonrandom, i.e., one for which the ciphertext
$\textbf{y}_n$ is uniquely determined by $\textbf{x}_n$ and the specific $\textbf{k}^{\mathsf{m}}$
used. From
her knowledge on the structure of a nonrandom cipher Eve could generate
the at most $2^{|K^{\mathsf{m}}|}$ possible sequences of $K_{n}'$  for any \textit{n} and determine
its probability distribution. For $p(K^{\mathsf{m}})=U$ , each of these $K_{n}'$  - sequence would
have the same probability $2^{-|K^{\mathsf{m}}|}$ under the normal nondegeneracy assumption of
exactly $2^{|K^{\mathsf{m}}|}$ number of $K'$ sequences.  The case of random ciphers is covered in
section IIC that treats classical noise and QKD generated keys.

\par From $p(K_{n}')$  one can determine Eve's maximum probability $p_{1}(K_{n}')$ of getting
the whole $K_{n}'$ correctly, which is just $2^{-\left|K^{\mathsf{m}}\right|}$
in the above typical situation.  From (1) Eve's mutual information per bit on $K_{n}'$ is at
least \begin{equation}\label{6}
 I_{E}(K_{n}')/n\geq1-H(K_{n}^{\mathsf{m}})/n
 \end{equation} The $\delta_{E}(K_{n}')$ is also large for typical
$n\gg|K^{\mathsf{m}}|$ similar to $I_{E}(K_{n}')/n$.  For a maximum length LFSR or more generally
nondegenerate ciphers, the probability of various consecutive subsequences $\tilde{K}'$ of $K_{n}'$
with
$|K_{n}'|<|K^{\mathsf{m}}|$ is given by $p_{1}(\tilde{K}')=2^{-|\tilde{K}'|}$ since such
$\tilde{K}'$ is uniformly distributed. Thus we have arrived at (2) in section I. However, different
subsequences of $K'$ are correlated
through $K^{\mathsf{m}}$ and no overall joint probability on any subset of ${\tilde{K}}'$
can be smaller than $2^{-\left|K^{\mathsf{m}}\right|}$.

\par It is clear from this description what the mechanism of the Shannon limit (1) is: however long
$X_n$ is for $n>|K^{\mathsf{m}}|$ , from $Y_n$ there are at most $2^{|K^{\mathsf{m}}|}$ possible
$X_n$ sequences from $\textbf{x}_n=\textbf{y}_n \oplus \textbf{k}_n$ or any injective encryption
map. Such a cipher is considered
adequate in standard cryptography for the protection of long data $X_n$ although it is far from
semantically secure, perhaps partly because
there is no alternative that does better information theoretically than what the key size
$|K^{\mathsf{m}}|$ allows which is always far less than $n=|X_n|$ . Perhaps it is partly due to (2),
and partly due to the practical complexity of $p(\tilde{K}')$ evaluation of more
general subsets $\tilde{K}'$. But see [20, section 5.5.3].

\subsection*{IIC. QKD Key Raw Security}

\par The case of classical noise key generation will be described first. In such cryptosystem where
noise is involved including randomized encryption systems, there is \textit{no longer} a fixed
observation random variable
$Y_{n}^{E}$ for Eve in contrast
to the case of the previous section on standard ciphers. In any key generation scheme, the user
Alice picks a random bit sequence $Z_{n'}$ of length $n'$ and transmits it to Bob via modulated
physical channel inputs, who could extract an error free (with high probability) bit sequence
$W_{n''}$ of length
$n^{\prime\prime}\leq n'$ from his observation. For example, this can be done with
an openly known error correcting code (ECC) so that both Alice and Bob know what
$W_{n''}$  is. Then a ``privacy amplification'' (PA) function $f_{\mathsf{PA}}$ is applied by
both to obtain the $n$ bit generated key
$K_n=f_{\mathsf{PA}}(W_{n''}(Z_{n'})),\,n<n''$.
One can combine the ECC and PA, to write $K_n=F(Z_{n'})$ for an openly known function $F$.  Eve
learns about $K_n$ through an
attack with some observed random variable $Y_{n'}^{E}$ that depends on $Z_{n'}$ via
$p(Y_{n'}^{E}|Z_{n'}),$ her probability of getting various  $\textbf{y}_{n}^{E}$ given the
possible $\textbf{z}_{n'}$. With $p(Z_{n'})=U$ and the known ECC and PA,  Eve then
obtains
$p(K_{n}|Y_{n^{\prime}}^{E}),$ to be called Eve's \textit{conditional probability distribution}
(CPD) on
her estimate of $K_n$. Note that a \textit{single number security criterion} is just a constraint on
Eve's possible CPD. The generation rate $r$ is usually taken to be $r = n/n'$, which is further
reduced in BB84 from basis matching.

\par
There is typically one complete $Y_{n^{\prime}}^{E}$ that Eve may observe in
classical-noise key generation and $Y_{n^{\prime}}^{E}$ can be drawn from a continuous alphabet, for
example when $p(Y_{n^{\prime}}^{E}|Z_{n^{\prime}})$ is given by that of an additive noise channel.
On the other hand, in QKD and KCQ different incompatible $p(Y_{n^{\prime}}^{E})$ may be obtained
from different incompatible quantum measurements, i.e., $p(Y_{n^{\prime}}^{E})$ depends on the
specific attack. In all cases the conditioning on different possible observations
$\textbf{y}_{n'}^{E}$
in $p(K_{n}|Y_{n'}^{E})$ is an added complication that does \textit{not} exist in standard ciphers.
We first
assume that $\textbf{y}_{n'}^{E}$ is fixed and order the resulting $p \left( \textbf{k}_{n} |
\textbf{y}_{n'}^E \right) \equiv p_i$ for $i \in \left\lbrace 1, \ldots,
N=2^{n}\right\rbrace\equiv \overline{1-N}$, $p_1\geq p_2\ldots\geq p_N$.
\par
The most common quantitative security criterion in classical-noise and QKD key generation is that
Eve's mutual information per bit $I_E(K_n)/n$ must be small, which cannot be satisfied in standard
cipher key expansion from (6). The \textit{question} is: how small is small enough for what
security? Note that this question does \textit{not} arise for the fresh keys in standard
cryptography where
the IT security can be presumed perfect. It is clear that Eve's maximum probability of getting the
whole key $K_n$ must be sufficiently small. With the distribution $p_1, p_2 = ...=p_n =
(1-p_1)/(N-1)$, it has been shown that [9, Section IIIB ]

\begin{flushleft}
Theorem 1:\\
For  $I_E/n\leq 2^{-l}$, there are CPD that give  $p_1\geq2^{-l}-\frac{1}{n2^{n}}$.

\end{flushleft}
Furthermore, there are CPD [9] that gives $p(\tilde{K}_m)\sim{\frac{n}{m}}p_1$ when $p_1 >> 1/N $
for \textit{m}-bit
subsets  $\tilde{K}_m$ of $K, m<n$.  Together these give the quantitative results (3), which gives
the operational significance of a $I_{E}/n$ guarantee.
\par
In many QKD security proofs Eve's possible $I_E/n$  is bounded for any possible attack she may
launch. Whatever one's notion of ``secure enough'' may be in a realistic application, the
experimental results thus far [16] that yields at best $l\sim10$ for $n>1,000$ is very
\textit{inadequate} as a security guarantee, because it does \textit{not} rule out the large
chance $10^{-3}$ of identifying the whole $1,000$ bit key with high probability. See section IVA for
a discussion on
the role of security proof in this connection. Thus, bounding $I_E/n$ insufficiently just leaves
open the possibility of a disastrous breach of
security. This situation cannot be expected to improve significantly in concrete realistic
protocols because from Theorem 1, decreasing the security parameter $I_E/n$ \textit{exponentially}
only leads to \textit{linear} increase in the effective number of random bits given by $l$. In this
connection, it may be observed that it is \textit{misleading} to consider ``exponentially
small'' as quantitatively adequate in key generation. See section IV.
\par
In the above $p_1$ for a given CPD we have suppressed the $\textbf{y}_{n^{\prime}}^{E}$ dependence.
The common criteria $I_E/n$ and $\delta_{E}$ are averages over all possible
$\textbf{y}_{n^{\prime}}^{E}$
for any given attack. As guarantee for each $\textbf{y}_{n^{\prime}}^{E}$, the Markov inequality
[24] for a
negative-valued random variable $X$ would need to be employed,
 \begin{equation}\label{7}
Pr[X\geq\epsilon] \leq E[X]/\epsilon
 \end{equation} where $E[X]$ is the average value of $X$.  Generally this would lead to the more
stringent requirement from an  $E[X]<\epsilon$ to $E[X]<\epsilon^{2}$ to guarantee $X<\epsilon$ with
probability $>1-\epsilon$, the latter probability requirement is especially appropriate when $X$ is
essentially a probability itself. Thus, by taking $I_E/n\sim p_1$ from (2) the actual full
guarantee would reduce the exponent $l$ by $\frac{1}{2}$, making it so much more difficult to
achieve a good value in practice.
\par The criterion of statistical distance ($L_1$-distance, variational distance, Kolmogorov
distance) $\delta_E\equiv\delta(P,U)$ between Eve's CPD $P=\{p_i\}$ and the uniform $U$ can be
used,

 \begin{equation}\label{8}
\delta(P,U)=\frac{1}{2}\sum^N_{i=1}\left|p_i -\frac{1}{N}\right|
 \end{equation} It is a direct consequence of the definition (8) that  [24, p.299]
 \begin{equation}\label{9}
\delta_E \leq \epsilon\quad \Longrightarrow \quad  p(\tilde{K}_m )\leq
\epsilon+\frac{1}{2^m}
 \end{equation}
where $\tilde{K}_m$ is any $m$-bit subset of $K_n$. As a numerical measure, $\delta_E$ suffers
exactly the same $p_1$ problem as $I_E/n.$ From the same distribution that gives Theorem 1 for
$p_1
=
2^{-l}$, one obtains [9, App B],
\begin{flushleft}
Theorem 2: \\
For $\delta_E=2^{-l}$, there are CPD that give $p_1= 2^{-l} + \frac{1}{N}$.
\end{flushleft}
\par Similar constructions give (9) with $\leq$ replaced by $=$, thus yielding (4) above.
Note that (9) gives the operational significance of a $\delta_{E}$ guarantee. From
Theorems 1 and 2 it follows that $K$ is only nearly uniform when $l\sim|K|$. It appears there is no
experimental protocol that has been quantified with $\delta_E$ or $d$. Theoretical estimates thus
far concentrate on $\epsilon = 10^{-5}$ for $d \leq \epsilon$ with various large $n$ [18]-[19]. Even
without using (7) for individual guarantee such $p_1$ is inadequate for
many purposes.
In particular, it is questionable to say a $10^{5}$ bit key $K$ has IT security with just $l\sim17$
as compared to the $l\sim 10^{5}$ for a really perfect key.
\par The raw security significance of $d \leq \epsilon$ is unknown as shown in the next
subsection. Thus, while QKD may in principle provide better $p_{1}(K)$ guarantee than conventional
key expansion it is much worse in practice and unlikely to significantly improve. This is because it
is difficult to capture the possible probability distribution
behavior with a mere single parameter value. A good $p_1$ guarantee, by itself or through
$I_{E}/n$ or $\delta_E$, is inadequate unless
$p_1=2^{-l}$ with $l\sim n$. This exponential-linear problem appears to be a major
quantitative stumbling block to physical cryptography.
\section* {III. composition security of the generated key}
\par In this section we will show that composition security has not been guaranteed at all in QKD
under $I_E/n$  or $\delta_E$ and their quantum counterparts. Recall that by the
\textit{composition security} of $K$ we mean its quantitative security level against attacks that
utilize information obtained both during the key generation process and during the actual
cryptogaphic use of $K$. We will discuss only a specific composition security scenario of
\textit{partial key leakage} (PKL) \textemdash whether it is possible for Eve to predict a future
bit exactly if part of $K$ is known to her.
\subsection* {IIIA. Conventional Cipher Composition Security}
\par We first review the case of a standard nonrandom cipher, for which the raw security of $K'$
was discussed in section IIB. For an additive stream cipher $\textbf{y}_n=\textbf{x}_n\oplus
\textbf{k}_n$ where $\textbf{k}_n$ is a specific value of $K_{n}'$ with distribution $p(K')$ on
$\overline{1-N}$, Eve would learn $m$ bits of $\textbf{k}_n$ from a KPA where she knows $m<n$ bits
of
$\textbf{x}_n$ and the openly observed value of $\textbf{y}_n$. From these $m$ bits of
$\textbf{k}_n$ Eve could try to determine the seedkey $K^{\mathsf{m}}$ to whatever degree possible
and with
such knowledge on $K^{\mathsf{m}}$ she could then get information on the other $n-m$ bits in
$\textbf{x}_n$
through $\textbf{y}_n$. Such KPA can often be launched in real world situations. It is
convenient to assume the $m$ known bits of $\textbf{x}_n$ are its
first $m$ bits.  For the situation where Eve knows nothing about $\textbf{x}_n$, i.e., $p(X_n)=U$,
the key $K'$ and hence $K^{\mathsf{m}}$ is totally hidden from observation of $Y_n$ alone since
$p(\textbf{k}_n|\textbf{y}_n)$ becomes a memoryless binary symmetric channel with crossover
probability $1/2$. That does not mean, of course, that the security of $X_n$ is good enough as
section IIB shows how it is limited by (1). It is possible that Eve knows something about $X_n$ so
that to her $p(X_n)\neq U$ but she does not know any bit in $\textbf{x}_n$
for sure.  We will not discuss such scenario of ``statistical attack''  [13],[14] in this paper and
focus on just KPA.

\par  For nondegenerate nonrandom ciphers [13],[14] there is a one-to-one mapping between
$K^{\mathsf{m}}$ and
$\{(X_{m'}, Y_{m'})|m'\,\in\, \overline{1-|K^{\mathsf{m}}|} \}$, the pairs of
$|K^{\mathsf{m}}|$
consecutive
data bits in $X_n$ and the corresponding output bits in $Y_n$.  This includes block ciphers
and their stream cipher modes of operation. Thus, in a KPA with $m=|K^{\mathsf{m}}|$ the key
$K^{\mathsf{m}}$
can be uniquely determined and the rest $n-m$ bits in $X_n$ are totally compromised information
theoretically. This is the situation in conventional symmetric-key ciphers such as AES where
security depends exclusively on the complexity of finding $K^{\mathsf{m}}$ from $\{(X_{m'},
Y_{m'})\}$. Indeed, such weak composition security is a manifestation of the weak $I_{E}/n$ raw
security from
the Shannon Limit (6). As will be seen in the next subsection IIIB, it is removed (classically) by
a strong IT guarantee on the raw security. Thus, the composition security situation of classical
key expansion is similar to both the raw and composition security of asymmetrical key ciphers such
as RSA whose
security depends on the complexity of factoring large integers.

\par Generally, we have the PKL problem of the extent to which knowledge on one part of the key $K$
would reveal about another part, all through the probability distribution $p(K)$ itself without any
further information as in the case of the above KPA. This problem does not exist for a
key $K$ with $p(K)=U$, not to mention a perfect key $K^{\mathsf{p}}$. The security
against PKL cannot be guaranteed by  $p_1$ alone, but it was thought that ``exponentially small''
$I_E/n$ and $\delta_E$ would be sufficient even in the context of QKD with Eve holding onto her
probe with quantum memory. In the rest of this section III we will show that is the
case in a purely classical situation but the presence of quantum effect, while allowing key
generation with small $I_E/n$ and $\delta_E$, also \textit{takes away} the composition security
guarantee that obtains in a purely classical scenario.

\par Before proceeding, it may be mentioned that the possibility of IT security on the key against
KPA is not ruled out for degenerate ciphers, meaningful versions of which can be developed for
classical ciphers with randomized encryption [13]. That is the subject for future
detailed treatment.

\subsection*{ III B. Composition Security Under $I_{E}$ and $\delta_{E}$ }

\par In this subsection we develop the composition security significance of an $I_E$ and an
$\delta_E$ guarantee on a key $K$ with distribution $p(K)$ to Eve. This applies to any classical
protocol directly and also to a quantum protocol through reduction of the quantum security criterion
guarantee. The quantum case would be handled in the
next subsection.

\par For a symmetric-key conventional cipher such as AES under KPA or statistical attack, Eve would
obtain in general information on $p(K)$ distributed through the whole $K$. We consider the specific
case where an m-bit subsequence $K_m$ of an n-bit $K$ is known exactly to Eve, for example obtained
from an m-bit KPA for $K$ used as one-time pad, and ask whether any other bit in $K$ would be
revealed with a significant probability from such knowledge. It is clear that a $p_{1}(K)$ guarantee
does nothing for this problem unless it is very close to $2^{-n}$, because it applies to just
one $\textbf{x}_n$ and it says little about correlations among the $n$ bits of $K$ which is the
matter of concern here. On the other hand,
since $I_E$ and $\delta_E$ are themselves already constraints on the whole $p(K)$, it may be
expected a sufficiently small value would lead to good composition security in this case. The
question is \textit{how} small.

\par The following result shows that a linear leak of information is possible under $I_E$ in the
absence of any quantum effect.
\begin{flushleft}
Theorem 3:
\par With $I_E/n = \epsilon$ for any $0 < \epsilon \leq 1 $ there are $p(K)$ for
which Eve knows one additional bit from knowing $\lceil 1/\epsilon \rceil$ number of them in an
$n$-bit $K$. Equivalently, for such $p(K)$ a fraction $\epsilon$ of the bits of $K$ can be
determined from the rest of $K$.
\end{flushleft}

\par The proof can be obtained from the following simple construction. Let $P(\textbf{k}_n) =
P(k_{1},...,
k_{n}) = 2^{-n}$ for $n$ bits of an ($n+1$)-bit $K$, and let $k_{n+1} = f(\textbf{k}_n)$
where $f$ is a known deterministic Boolean function of $\textbf{k}_n$. Then $P(\textbf{k}_{n+1}) =
P(k_{n+1}|\textbf{k}_{n})P(\textbf{k}_n) = 2^{-n}$ from which it follows that $I_{E}/(n+1) =
1/(n+1)$. Either by extension to $m > 1$ other bits determined by $\textbf{k}_n$ or by forming a
product distribution, the theorem follows.

\par Thus under a KPA with $I_{E}/n \leq 2^{-l}$, one bit may be leaked for every $2^{l}$ known
bits. This is perhaps tolerable in some applications when $l \geq 10$, although there is still the
issue of the distribution of such leaked bits in $K$. Unfortunately, the quantum
situation is much \textit{worse} as discussed in subsection IIIC with respect to the corresponding
accessible information.

\par The situation for $\delta_{E}$ is much more favorable than $I_{E}$. It is easy to show by
simple counting that no deterministic bit of $K$ can be leaked this way when $\delta_{E} < 1/2$. It
can be shown that for $\delta_{E} \leq \epsilon$ Eve could not achieve even a probability of knowing
a bit better than $\epsilon + 1/2 $ compared to $1/2$ from pure guessing. We will not dwell
on the composition security of a $\delta_E$ classical guarantee with no quantum probe that has been
held in quantum memory, as it appears entirely adequate. On the other hand, as will be discussed in
the next subsection IIIC there is no known guarantee in the case when quantum memory is available
for a related quantum criterion $d$. A
contrary claim has been repeatedly made in the literature [25] including a recent broad review of
QKD security [5]. The \textit{error} in such a claim can be traced to an incorrect inference on the
meaning
of the classical statistical distance to be presently discussed.

\par It was suggested that between two distributions $P,Q$ for two random variables $X$ and $X'$
over the same range $\chi$ of $N$ elements, the statistical distance \begin{equation}\label{10}
\delta(P,Q) \equiv \frac{1}{2} \sum_{\textbf{x} \in \chi} \left|P(\textbf{x}) - Q(\textbf{x})\right|
\end{equation} ``can be interpreted as the probability that two
random experiments described by $P$ and $Q$ respectively, are different'' [11],[17] , an
interpretation repeated in refs. [12], [28]. The justification for the interpretation is given by
lemma 1 in refs [11], [28] which states that for any two distributions
$P$ and $Q$ for $X$ and $X'$ there exists a joint distribution $P_{XX^{\prime}}$ that gives $P,Q$
as marginals with
\begin{equation}\label{11}
Pr[X\neq X^{\prime}] = \delta(P,Q).
\end{equation}
However, to the extent it makes sense to talk about such a joint distribution, the interpretation
would obtain only if
``there exists'' is replaced by ``for every''. This is because since there is no knowledge on such
joint distribution, one
cannot assume the most favorable case via ``there exists'' for security guarantee or for general
interpretation. Indeed, it is
not clear at all what realistic meaning can be given or claimed for the realization of such a joint
distribution, other than the
independent case $P_{XX^{\prime}}=P\cdot Q$. This independent case is the appropriate one to
consider since one is just
comparing
two distributions $P$ and $Q$ with $\delta(P,Q)$. In such case, even if both $P$ and $Q$ are the
same
uniform distribution so that $\delta(P,Q)= 0$,
we have $Pr[X\neq X^{\prime}]=1-\frac{1}{N}$  and the two sides of (11) are almost as far apart as
it could be since both must be between $0$ and $1$. This is also a counter-example to the
interpretation. As a matter of fact, whether (11) holds is irrelevant to how close P and Q are
according to $\delta (P,Q).$

\par Furthermore, instead of (11) the following equation (12) is a consequence of the
interpretation, \begin{equation}\label{12} P(\textbf{x}) = (1-\epsilon)Q(\textbf{x}) + \epsilon
P'(\textbf{x}) \end{equation} where
$P'$ is a probability distribution on $\chi$. Indeed (12) may be taken as the mathematical
representation of the interpretation, apart from possible ``partition ensemble fallacy'' which we
would not discuss since there is no need. However, (12) cannot be true when $\delta(P,Q)=\epsilon$
because that occurs if and only if $\delta(P', Q) = 1$ which in turn holds if and only if $P'$
and $Q$ are never both nonzero on the same $\textbf{x}$. The latter never occurs when $\chi$ is
taken to be
the
common range of $P$ and $Q$ as indicated. Thus, not only the interpretation is not proven, it has
nothing to do with (11) and \textit{cannot} even be true. One can also see this from the
immediate fact that $P \neq Q$ for sure whenever $\delta(P,Q)=\epsilon>0$.

\subsection*{IIIC. Composition Security in QKD and the Criterion d }

\par In the quantum case Holevo's bound is often used to bound the accessible information $I_E$ Eve
may obtain via some quantum measurement in the key generation process. In the presence of enduring
quantum memory, however, Eve may utilize the bit knowledge she obtained on parts of $K$ via KPA
in conjunction with the quantum probe still in her possession to get at the rest of $K$. In this
quantum case there is the additional issue of lockable information [12], that a random
variable side information $S$ may reveal to Eve more than $H(S)$ bits of information on $K$ which is
impossible classically. Without the quantum probe there would be at most $H(S)$ bits of classical
information which is already reflected in $H_E(K)$ for the PKL problem. In [29], it was suggested
that if Eve's optimal
mutual information on $K$ from a quantum measurement, called the accessible information
$I_\mathrm{acc}$, is exponentially small in $n$ for large $n$, then the $n$-bit $K$ is composition
secure according to their quantitative definition.  While the mathematical result in [29]
is correct, it was pointed out in [12] via a counter-example with one-time pad use of $K$
that the result does not have the interpretation given in [29] to guarantee their
composition security. In particular, it was shown that for $I_{E}/n \sim 2^{-l}$, each $l$
bits of knowledge on $K$ may yield another bit. For $I_{E}/n \leq \epsilon$ the fraction of bit
leakage may thus go up from $\epsilon$ to $log \frac{1}{\epsilon}$, an exponential increase from the
case where no quantum memory is available. For $l = 10$, the possible leak thus increases from
$0.1\%$ to $10\%$.

\par The following remedy was suggested [11],[17] by using the criterion (5). Let
$\rho_E^\textbf{k}$ be the state in Eve's possession conditioned on a generated key value
$\textbf{k}$, and let

 \begin{equation}\label{13}
\rho_U \equiv\frac {1}{|K|} \sum_\textbf{k} |\textbf{k}\rangle \langle \textbf{k}|
\end {equation} be the completely mixed uniform state on the $|K|=2^n$ orthonormal
$\ket{\textbf{k}}$'s. It is assumed that the a priori probability of $K$ to Eve before she
measures on her probe is uniform. Then \emph{the security criterion} $d$ is the trace distance
 \begin{equation}\label{14}
d\equiv \frac{1}{2} \norm{\rho_{KE}-\rho_U \otimes \rho_E}_1
\end {equation}
where
 \begin{equation}\label{15}
\rho_{KE}\equiv \frac {1}{|K|}\sum_\textbf{k} \ket{\textbf{k}}\bra{\textbf{k}} \otimes
\rho_E^\textbf{k}
\end {equation}
and
 \begin{equation}\label{16}
\rho_E \equiv \frac {1}{|K|} \sum_\textbf{k} \rho_E^\textbf{k}.
\end {equation} The trace distance $\|\rho-\sigma\|_1$ between two states is related to the
classical statistical distance $\delta(P,Q)$ between two probability distributions as follows [11].
For any POVM or von Neumann measurement made on $\rho$ and
$\sigma$ with resulting distribution $P$ and $Q$, \begin{equation}\label{17}\|\rho -\sigma\|_1 \leq
\epsilon \quad \Longrightarrow \quad \delta(P,Q) \leq \epsilon\end{equation} Using (17) and (11) it
is
concluded [11, p. 414], [17, Prop 2.1.1] that when $d \leq \epsilon$ the key is
\textit{$\epsilon$-secure}: with probability $p \geq
1-\epsilon$ the real and the ideal situation of perfect security can be considered identical, where
the ideal situation is one where $K$ is replaced by a uniformly distributed random variable $U$
which is independent of $\rho_E^{\textbf{k}}$. This statement is repeatedly made [25]
and
provides the following two very desirable consequences to supply both raw and composition
security significance. Under $d \leq \epsilon$, with probability $p
\geq 1 - \epsilon$ the key $K$ is universally composable (or at least so for partial key leakage)
and it is the same as the uniform $U$ to Eve for the raw security apart from composition. Note that
similar to the invalidity of (12), $d = \epsilon$
does \textit{not} imply \begin{equation}\label{18} \rho_{KE} = (1-\epsilon) \rho_U \otimes \rho_E +
\epsilon
\sigma_{KE} \end{equation}for some joint density
operator $\sigma_{KE}$. Equation (18) may lend itself to the above incorrect interpretation.
Specific counter examples and further discussion of (18) can be found in ref [30].

\par To see how (5)
does not give an $\epsilon$-secure key and how $\delta_E$ enters, we now trace the steps of the
above incorrect derivation [11]. Let $P_{\textbf{y}}^{\textbf{k}}\equiv P(\textbf{y}|\textbf{k})$ be
Eve's
probability distribution on her measurement result $\textbf{y}$ conditioned on an actual generated
$\textbf{k}$, i.e., through $\rho_{E}^{\textbf{k}}$, with $P_{\textbf{y}} = P(\textbf{y})$ the
distribution obtained through $\rho_E$. For $d \leq \epsilon$, (17) and (14) imply, with Eve's a
priori distribution on $\textbf{k}$ given by $U,$
\begin{equation}\label{19}
\delta(P_{\textbf{y}}^{\textbf{k}}U_{\textbf{k}}, P_{\textbf{y}}U_{\textbf{k}}) \leq \epsilon
\end{equation} Under the incorrect interpretation, this implies $P_{\textbf{y}}^{\textbf{k}} =
P_{\textbf{y}}$ independent of $\textbf{k}$ with probability $\geq 1 - \epsilon,$ thus Eve's
corresponding CPD $P(\textbf{k}|\textbf{y})$ is equal to $U$ also with probability $\geq 1 -
\epsilon.$ Not only
this conclusion does not follow unless (11) is true, one can see that
$P(\textbf{k}|\textbf{y})\neq U$ for sure as long as $\rho_{E}^{\textbf{k}}$
or $P_{\textbf{y}}^{\textbf{k}}$ carries any $\textbf{k}$-dependence, i.e., when there exists
$\textbf{k}_1 \neq \textbf{k}_2$ with $P_{\textbf{y}}^{\textbf{k}_1} \neq
P_{\textbf{y}}^{\textbf{k}_2}$ for a given $\textbf{y}.$ This is because
$P_{\textbf{y}}^{\textbf{k}}$ then depends on $\textbf{k}$ and thus Eve cannot have $U$ as her CPD
for the given $\textbf{y}.$

\par To derive the raw security meaning of $d\leq\epsilon,$ we first observe
that since Bob and Eve perform their ``local'' operations separately, the criterion
$d$ is \emph{exactly} equivalent to \begin{equation}\label{20}
d = \mrm{E}_\textbf{k} [\norm{\rho_E^\textbf{k}-\rho_E}_1]
\end {equation} and (5) is just a condition on the $\rho_{E}^{\textbf{k}}$.
With $\mrm{E}_\textbf{k}$ the average over the $2^n$ possible values of $K$, equality of the right
hand sides
of (14) and (20) follows from lemma 2 of ref. [11]
directly. The right-hand side of (20) is actually one of the criteria proposed in [29].

\par Apart from an increase in $\epsilon$ from a Markov inequality guarantee for individual
$\textbf{k}$, (20) implies
\begin{equation}\label{21}
 \norm{ \rho_{E}^{\textbf{k}} - \rho_{E} }_{1} \quad \leq \epsilon
\end{equation} From (21), (17), and (10) it follows that given $n'$-bit $\textbf{y}_{n'}^E ,$ Eve's
CPD $P(\textbf{k}|\textbf{y})$ satisfies

\begin{equation}\label{22}
 \left| P(\textbf{k}|\textbf{y})-U_{\textbf{k}}\right| \leq \epsilon \cdot
U_{\textbf{k}}/P_{\textbf{y}}.
\end{equation} The best guarantee from (22) on the smallest $P(\textbf{k}|\textbf{y})$ over all
possible $n'$-bit $\textbf{y}_{n'}$ is $P_{\textbf{y}_{n'}} = U_{\textbf{y}_{n'}},$ i.e., the
minimax of $P_{\textbf{y}}$ over $\textbf{y}$ and $P_{\textbf{y}}$ is obtained by $P_{\textbf{y}} =
U_{\textbf{y}}.$ Exactly similar results are obtained for subsequences $\textbf{y}_{m'}$ of
$\textbf{y}_{n'}$ corresponding to (9). Thus, (22) would reduce to $\delta(P,U) = \delta_{E} \leq
\epsilon$ when the number of possible $\textbf{y}$'s is $N$. With key sifting, error correction and
privacy amplification, the number of such possible $\textbf{y}$ is much larger and so
$U_{\textbf{k}}/P_{\textbf{y}}$ is a very large number that would render the guarantee (22)
useless. Thus, (21)-(22) does \textit{not} turn into a useful $\delta_{E} \leq \epsilon$  guarantee
on Eve's CPD. This also means (5) implies no composition security guarantee through $\delta_{E}$
for the case of no quantum memory.

\par An $\epsilon$-secure key is evidently ``universally composable'' as concluded previously.
Since $d \leq \epsilon $ does not imply the key is $\epsilon$-secure, the composability problem
remains in the presence of quantum memory. It is yet \textit{not} known what the level of PKL
leakage
may be under the $d\leq \epsilon$ guarantees, whether leakage similar to the accessible information
case is ruled out.

\par Since $\delta(P,Q) > 0$ implies $P\neq Q$ for sure [31], we have the following
situation
\begin{description}\item[(A)] Under $d = \epsilon$, Eve's probability
distribution $p(K)$ is \textit{not} the uniform $U$ for sure.
\end{description}
which can be compared to the following claim in the literature [25]
\begin{description}
\item[(A')] Under $d = \epsilon$, Eve's probability distribution $p(K)$ is $U$ with probability
$1-\epsilon$.
\end{description} Note the huge difference between between (A) and (A'), the latter is currently
used
to justify the IT security guarantee of a QKD key. On the other hand, (A') is actually impossible
from
(A), and the probabilistic or operational significance of $d\leq \epsilon$ for both the quantitative
raw and composition security are unknown.

\section*{IV. Importance of Security Proof and Numerical Values}

\par Physical cryptography raises new conceptual issues in addition to the already subtle ones in
both conventional symmetric-key and asymmetric key cryptography. In order to fully assess the
significance of the above results for actual security guarantee, we will discuss some such
conceptual issues in this section.

\subsection*{IVA. Security Proof in Physical Cryptography}

\par We first observe that, in contrast to almost all problems in physics and most in engineering,
a guarantee of cryptographic security \textit{cannot} be obtained by experiments which could show a
task is well carried out but not something general is impossible. An experiment can implement a
specific attack and show that it does not work, but one cannot implement all possible attacks.

\par All proofs are based on reasoning on specific givens, in this case the  mathematical model of
the physical cryptosystem must be valid for the actual situation if the proof is to mean what it
says in a real application. In standard or conventional cryptography where purely mathematical
relations constitute the entire security mechanism, there is already a problem on the realistic
features of an operating cryptosystem that cannot be incoporated in a general mathematical
representation and must be treated on an individual ad hoc basis, such as the case of the RSA
timing attack. In a physical
cryptosystem involving either classical noise sources or quantum effects, the actual mathematical
representation is a major issue due to the presence of other interfering physical effects that may
play a
crucial role in the actual cryptographic security. In particular, quantum information is an unusual
area in physics where very small disturbance can lead to major consequence. In BB84 type QKD there
is
a serious problem of system and device modeling \textit{at the time of use}, see, e.g., ref
[32]-[35], which arises from the single-photon nature of the signal. In particular, a device
imperfection can entirely compromise
the security of a BB84 protocol [34]-[35]. The issue here is not that the device imperfection
cannot be removed, but rather how many such \textit{undiscovered} loopholes there are in practice.
However, in this paper we do not deal with these issues
but only with the fundamental quantitative security assuming the physical model is exactly correct.

\par The excitement of physical cryptography and particularly QKD is mainly derived from the belief
that unconditional information-theoretic security is possible for generating fresh keys which can
be \textit{proved} mathematically given a model. This is in sharp contrast to conventional
cryptography, in which asymmetric key crypto-system has only complexity based security the strength
of which is further based on \textit{unproved} though widely accepted assumptions on the
difficulties of various mathematical problems. For symmetric-key conventional ciphers against KPA,
the design is even more of ``an art'', with security based on less widely shared beliefs in the
problem complexity of various attack algorithmss. In QKD, ``\textit{unconditional
security}'' means all possibilities
of an
attacker gaining more ``information'' than a designed level is \textit{ruled out} except for a small
probability which is itself a design parameter. That is, claim (a) in the Introduction of this
paper is maintained.

\par In contrast to a perfect key $K^{\mathsf{p}}$ the QKD generated key $K$ can never be perfect
because
Eve could always obtain some ``information'' by an attack during the physical key
generation process. The crucial questions are then what operational meaning the various security
criteria and proofs have. These questions
are already subtle ones in conventional cryptography, see for example the dispute described in ref
[36] on public-key systems and the complaint on lack of proper security foundation in symmetric-key
ciphers [20]. In physical cryptography such questions are much more acute while similar security
situation has not arisen previously in any real cipher. These problems do arise in a more
restricted manner (no quantum memory) in classical-noise key generation which, however, has never
found
publicly known actual deployment, and the criterion of $I_{E}/n$ was employed without any discussion
on its adequacy in cryptographic context [1]-[3].

\par The various probabilities one can obtain from a mathematical model have a clear empirical or
operational meaning, in the same sense that probability in quantum physics or communication
engineering has empirical meaning. However, various theoretical constructs such as $I_E$ and
$\delta_E$
do not automatically have
the meaning that would ensure whatever security we may desire in an application. They are really
no more than \textit{mere constraints} on the possible distribution $p(K)$ Eve may obtain in an
attack and need to be transformed into operational guarantees as done in this paper. In particular,
it is misleading to
claim that the system is secure if $\epsilon$ can be made exponentially small in $| K |$, as the
following shows.

\begin{table*}
\begin{tabular}{|cc|c|c|c|c|l}
\cline{1-5}
\multicolumn{2}{|c|}{Criteria} & $p_{1}(K) \leq \epsilon$ & $I_{E}/n \leq \epsilon$ & $\delta_{E}
\leq \epsilon$  \\ \cline{1-5}
\multicolumn{2}{|c|}{\multirow{2}{*}{Raw Security}} &
leak of $K$ & $p_{1}(K) \sim \epsilon$,&  $p_{1}(K) = \epsilon + \frac{1}{N}$,
\\ & & with probability $\epsilon$ & $p_{1}(\tilde{K}) \sim
\frac{\left|K\right|}{\left|\tilde{K}\right|}\epsilon$ & $p_{1}(\tilde{K}) = \epsilon +
\frac{1}{2^{ \left| \tilde{K} \right| }}$
\\ \cline{1-5}
\multicolumn{1}{|c|}{\multirow{1}{*}{Composition}} &
\multicolumn{1}{|c|}{\multirow{1}{*}{no quantum}} & \multicolumn{1}{|c|}{\multirow{2}{*}{$f
\sim 1 - \epsilon$}} & \multicolumn{1}{|c|}{\multirow{2}{*}{$f \sim \epsilon$}} &
\multicolumn{1}{|c|}{\multirow{2}{*}{$f \sim 0$}}  \\
\multicolumn{1}{|c|}{\multirow{1}{*}{Security \textbf{\textemdash\textemdash}}}
  &
\multicolumn{1}{|c|}{\multirow{1}{*}{memory}} & & &   \\ \cline{2-5}
\multicolumn{1}{|c|}{\multirow{1}{*}{fraction $f$ of $K$}}
&\multicolumn{1}{|c|}{\multirow{1}{*}{with
quantum}} & \multicolumn{1}{|c|}{\multirow{2}{*}{$ f \geq 1 - \epsilon $}} &
\multicolumn{1}{|c|}{\multirow{2}{*}{$f \geq
log\frac{1}{\epsilon} $}} & \multicolumn{1}{|c|}{\multirow{2}{*}{$f \sim ?$}}
\\
\multicolumn{1}{|c|}{\multirow{1}{*}{revealed in PKL}}&
\multicolumn{1}{|c|}{\multirow{1}{*}{memory}}  & & &
\\ \cline{1-5}

\end{tabular}

\caption{Quantitative Security in QKD}
\end{table*}

\par For example, for a 1,000 bit $K$, if $I_{E}/n \lesssim 2^{-20}$ which is an ``exponentially
small'' number to many, one may thus claim $K$ is ``secure''. From Theorem 1 it is not ruled out
that Eve may
identify
$K$ with a probability $2^{-20}$. Since $I_{E}/n$ is obtained under average over all possible key
values, from (7) and the surrounding discussion the final security  guarantee then becomes: with a
probability $\geq 1 -
2^{-10}$, Eve has $I_{E}/n \leq 2^{-10}$ and thus she may not be able to get the whole $K$ from her
measurement with
probability more than $2^{-10}$. This is a weaker security guarantee than the simple statement that
for sure Eve could not get $K$ with probability more than $p_1 =  2^{-10}$ from her
measurement. Since $|K| = 10^{3} >>10$, such guarantee clearly does \textit{not} rule
out with a small enough overall probability of a  disastrous breach of security that Eve determines
the whole $K$ with probability $2^{-10} \sim 10^{-3}$ from her measurement result alone without any
further use
context on $K$. Even for
$p_1 = 2^{-100}$, one may ask in \textit{what} sense such a $K$ is near a perfect 1,000 bit
$K^{\mathsf{p}}$,
particularly in view of all the other subset breach probabilities given in (3).

\par It is the \textit{role} of a security proof to rule out such disastrous breach of security
\textbf{\textemdash\textemdash} here it is Eve being able to identify $K$ at a probability $p_1
\sim 10^{-3}$ from her measured
result  \textbf{\textemdash\textemdash} by making it very
unlikely if not impossible, certainly not at a probability $\sim 10^{-3}$. Security breach with
probability $\sim 10^{-3}$ is only a possibility, whether Eve can actually do it depends on what
her distribution $p(K)$ is, which is obtained via her specific attack that gives her $p(K)$. A QKD
guarantee in terms of accessible information shows no $I_{E}/n$ can
exceed a designed level $\epsilon$ when averaged over specific $\textbf{k}$. Such guarantee leaves
open the
above possibility that cannot be further averaged out \textbf{\textemdash}
Eve knows her $p(K)$ which is fixed by her attack. Indeed, the possible large leakage from
accessible information guarantee in composition security when Eve has quantum memory as given in
ref [12] is \textit{exactly} the same in this regard \textbf{\textemdash} it shows a serious
compromise of security is not
\textit{ruled out} under the security condition given in [29].

\par A security proof via $\delta_{E} \leq \epsilon$ or $d \leq \epsilon$ is exactly of the same
nature, and $d\leq\epsilon$ is adequate only if one uses the mistaken interpretation (A') in
subsection IIIC instead of the correct
(A). Unless $I_{E}/n$ or $\delta_{E}$ is close to $2^{-|K|}$, the QKD generated
$K$ is very far from a perfect $K^{\mathsf{p}}$ while a security proof of $d \leq \epsilon$ has
uncertain quantitative significance in terms of empirically meaningful probabilities. We will now
discuss the numerical situation further specifically.

\subsection*{IVB. Actual Quantitative Guarantee}

\par We summarize the status of the quantitative security  guarantee situation in Table 1. Note
that $d \leq \epsilon$ is not listed because it has no clear security significance. The
security parameter is $\epsilon$ and smaller $\epsilon$ means the system is more secure. Recall
that raw security measures the information Eve has just from her attack before the key $K$ is used,
and composition security in this case refers only to the fraction $f$ of deterministic bits that
Eve could get on $K$ from knowing the rest of $K$ as in a KPA. The $\tilde{K}$ are subsequences of
the $n$-bit $K$, $N = 2^n$. All the filled entries in the table other than ``$f\sim$?'' are
worst case leaks except for
the composition security  under $I_{E}/n \leq \epsilon$. In that case the leak of $f \sim \epsilon$
would also occur in some probabalistic form other than deterministic bits of $K$ in the case of no
quantum memory. With quantum memory $f
\sim log \frac{1}{\epsilon}$ has not been shown to be the worst scenario. It is important to
observe that the fraction $f$ of bits leaked can be distributed in any fashion and not just
uniformly in $K$. Thus, there could be a \textit{serious} security breach even when $f$ is very
small. This shows the importance of semantic security.

\par In current experimental scenarios the final generated key in a single cycle, or round of QKD
after error correction and privacy amplification, could have $|K|$ in thousands of bits or more.
The necessary message authentication shared secret key $K^{a}$ that is needed to create a ``public
channel'' in BB84 type
protocols has never been explicitly integrated into the protocol and accounted for. It is reasonable
to assume $|K^{a}| \sim 100$ for each round is to be used with one of the current message
authentication code, since $|K^{a}| \gtrsim 40$ is typically used for many such  codes. Compared
to the alternative use of $K^a$ as the seed key $K^{\mathsf{m}}$ in a conventional cipher, it is
clear that
but for KPA there is little point in using a QKD generated $K$ from the viewpoint of security
guarantee. When the input data can be
assumed uniformly random to Eve, such conventional cipher $K'$ would give better protection than
the QKD keys that can readily be generated in the foreseeable future. This is evident from the fact
that even without taking (7) into account, the best
current $I_{E}/n \sim 2^{-10}$[16] and $\delta_{E}$ or $d$ has apparently
not
been used in an actual experimental system while theoretical estimates give $d = 10^{-5} \sim
2^{-17}$ [18]-[19]. In all these cases $n$ is $10^3$ or much larger. In the literature such
numerical evaluation was never compared to the benchmark of a perfect $K^{\mathsf{p}}.$

\par One \textit{main problem} in this connection is that almost every relevant quantity is
``expontentially small'' here. If one takes that to mean $2^{-\lambda n}$, $0<\lambda \leq 1$, and
$|K| = n$, it all depends on how large $\lambda$ is. Indeed $n$ is not ``asymptotic'' either in a
real protocol. Thus, the actual security depends on the precise numerical values of the system
parameter and one \textit{cannot} capture the situation by a vague qualitative remark. Other than
$p_{1}(K)$
for identifying the whole $K$, under (5) for $\epsilon = 2^{-l}$ any large subset $\tilde{K}'$ of
$K$ may still be determined with a much larger probability $2^{-l}$ than the uniform $|\tilde{K}|$
one of $2^{-|\tilde{K}|}$. It is clear that the incorrect statement (A') in section IIIC is sorely
needed for good security guarantee, but (A) shows in a strong way that (A') is forever unachievable.

\section*{V Conclusion and Outlook}

\par In this paper we have seen that realistic QKD generated keys have inadequate raw security and
no composition security guarantee. One may obtain much better raw security
probability guarantee with conventional ciphers. The perception to the contrary is due mainly to a
mistaken
interpretation of the security criterion $d \leq \epsilon,$ which has actually no clear operational
security
significance. One may view the logical/historical development on the information
theoretic security of the generated key as follows. The Shannon limit on conventional key
expansion leads
to very poor composition security under known plaintext attack. This composition security
predicament is rectified by QKD via the mutual information criterion when the attacker does not
possess good quantum memory, but the problem remains when she does. At the same time the raw
security guarantee worsens in concrete QKD protocols. Since $I_{E}/n \leq
\epsilon \sim 2 ^{-\lambda|K|}$ needs to go
down exponentially for linear bit improvement in the security of $K$, it does not appear promising
to try a brute-force experimental approach for increased security of either the raw or composition
kind. The criterion $\delta_{E} \leq \epsilon$ may be used in a classical noise protocol, it
provides good PKL security but has problems simliar to $I_{E}/n$ for raw security. It is
\textit{not}
clear how its quantum generalization may be developed and what its composition security would be in
the presence of quantum memory. In the absence of adequate guarantee on both raw and
composition security, QKD would \textit{lose} its main claim of merit over conventional cryptography
and would \textit{reduce} in practice to a mere ``art'' similar in many ways to the latter.

\par What can be done about it? The following four alternative routes may be
suggested:
\begin{enumerate}[label=(\roman{*})]
 \item At the expense of efficiency, it may be possible to improve security under (3)-(4) by
appropriate privacy amplification. However, privacy amplification cannot improve $p_1$ [9,section
IIID]. Due to the small $l$ that can be obtained, this does
not look promising for a real protocol to get $l$ (near) uniform bits in $K$ from $p_{1}
\sim 2^{-l}$ even if possible. The effective key generation rate would be reduced from $r$ to
$r'=rl/n$ for an n-bit $K$.
  \item One may use more efficient key generation schemes from the KCQ approach [9] other
than QKD
with intrusion level estimation. The possibility of obtaining adequate security with such approach
against all attacks can be explored.
\item
One may limit the security to just the more realistic attacks that can be launched with
foreseeable technology advance. This would rule out, in particular, joint attacks that involve
actual quantum entanglement over several or more subsystems. The situation may then be reduced to
that of
a wiretap channel [1] and one may genearte near-uniform $K$ with a nonvanishing final key rate [37].
\item One may limit the devices Eve possesses to more realistic ones. In particular,
this would exclude long and near-perfect quantum memory and help the composition security instantly.
Devices that
are totally free of
the many limits that have been around for a long time may also be excluded. This restriction is in
addition to and
independent of that in (iii), as it concerns with device realization rather than unknown
in-principle schematic realization although both can be brought under the general classification of
limited technology.

 \end{enumerate}

\par Given the subtle modeling question in physical cryptography and especially in BB84 type
protocols, it is not clear that (iii)-(iv)
entail any loss of true security in a real world application as compared to the inadequate levels
one may
obtain in an ideal model that allows Eve all the physical possibilities. Note that all
conventional cryptosystems are being currently deployed under equivalent assumptions to (iii)-(iv)
on unavailable alrogithms and computing power,
which are of a mathematical nature instead of physical ones. It is not clear why mathematical
presumptions are better than physical ones. One may argue the contrary in some situations. The clear
advantage of physical cryptography is that it is difficult to launch an attack or
to obtain just the ``ciphertext'', in sharp contrast to conventional cryptosystems. It is possible
that feature alone is enough to justify the deployment of physical cryptosystems in some
applications.

\section*{Acknowledgements}
\par This work was supported by AFOSR and DARPA.


\begin{thebibliography}{2}
\bibitem{aj} The term ``BB84 type protocols'' would be used to cover variants including Ekert
protocols that employ quantum entanglement explicitly. The key
feature of such protocols is that intrusion level estimation via public exchange is used to bring
about security.
\bibitem{ay} This is explicitly stated in
[11]-p.414, [17]-Section2.2.2, [26]-p.012332-5, [12]-p.140502-3, [18]-p.200501-2,  [27, p.02054-3],
[5, p.1310].
\bibitem{be} It follows immediately from definition that $\delta(P,Q)>0$ implies $P\neq Q$ for
sure. With $d = \epsilon$ one only gets $\delta(P,Q) \leq \epsilon$ from (17) and thus $\delta(P,Q)
= 0$ is not strictly ruled out. However, that appears to occur only in a highly degenerate and
poorly chosen attack.
\bibitem{bk} This is different from the situation of [27] which uses the criterion $d \leq
\epsilon,$ and will be further discussed elsewhere.
\end{thebibliography}
\end{document}